# Price Updating in Combinatorial Prediction Markets with Bayesian Networks

**David M. Pennock**
Yahoo! Research New York
111 West 40th Street, 17th floor
New York, NY 10018

**Lirong Xia**[*]
Department of Computer Science
Duke University
Durham, NC 27708, USA

## Abstract

To overcome the #P-hardness of computing/updating prices in *logarithm market scoring rule-based (LMSR-based)* combinatorial prediction markets, Chen et al. [5] recently used a simple Bayesian network to represent the prices of securities in combinatorial prediction markets for tournaments, and showed that two types of popular securities are *structure preserving*. In this paper, we significantly extend this idea by employing Bayesian networks in general combinatorial prediction markets. We reveal a very natural connection between LMSR-based combinatorial prediction markets and probabilistic belief aggregation, which leads to a complete characterization of all structure preserving securities for *decomposable* network structures. Notably, the main results by Chen et al. [5] are corollaries of our characterization. We then prove that in order for a very basic set of securities to be structure preserving, the graph of the Bayesian network must be decomposable. We also discuss some approximation techniques for securities that are not structure preserving.

## 1 Introduction

In a prediction market, agents trade securities about the outcome of an uncertain event, for example, "if event $E$ happens, this security pays off \$1". If $E$ does happen, the agent receives \$1 for every share of the security owned; if $E$ does not happen, the agent gets nothing. The price of a security reflects a collective value for "\$1 if $E$", or a group-wide or consensus probability of the event $E$. The idea is to harness market efficiency and leverage agents' incentives to earn money in order to price events that might not otherwise be traded. In a prediction market, the primary goal is price discovery and thus information aggregation, not finding gains from trade. The *Iowa Electronic Market* and *Intrade* are two examples of real-world prediction markets with a long history of tested results [1, 2]. See Chen and Pennock [7] for a recent survey of prediction mechanisms.

In this paper, we focus on prediction market mechanisms that employ a central *market maker* that determines the prices of securities algorithmically based on a *cost function* [6]. At any time, an interested agent can query the market maker for the price of a security and can either "take it or leave it": that is, decide to buy or sell some shares at the quoted price, or do nothing. After each (infinitesimal) trade, the market maker updates the prices of all securities. For example, suppose there is a prediction market on a Duke basketball game, and the current price for the security "Duke wins" is \$ $0.8$. If a risk-neutral agent believes that Duke will win with $0.9$ probability, then she has an incentive to buy some shares of the security, because her expected profit for holding one share is $0.9 - 0.8 = 0.1$. If she buys some shares of the security, then its price will go up; on the other hand, if she sells some shares of the security (in this case, equivalent to buying shares of Duke's opponent), then its price will go down. See Section 2 for more details.

So far most work has been focusing on the prediction markets based on the cost function that corresponds to the *logarithm market scoring rule (LMSR)* [4, 5, 8, 16, 17]. Pricing securities in LMSR-based prediction markets by directly using the cost function takes time that is polynomial in the number of outcomes. Therefore, it works well if the number of outcomes is not too large. However, in many real-life situations the number of outcomes is exponentially large and has a combinatorial structure. Such situations are called *combinatorial prediction markets* [4, 5, 11, 16, 17]. For example, in the NCAA men's basketball tournament, there are $64$ teams and therefore $63$ matches in total to predict. Each match can be seen as a binary variable. It follows that the prediction market for this tournament has $2^{63} \approx 9.2 \times 10^{18}$ outcomes. Therefore, in such situations, computing and updating the prices by directly using the cost function becomes computationally intractable. In fact,

---

[*]Part of this work was conducted at Yahoo! Research.

pricing LMSR-based combinatorial prediction markets is #P-hard [4].

In LMSR-based prediction markets, at any time the prices of the securities that correspond to disjoint, exhaustive outcomes sum up to 1. Therefore, the market prices can be seen as a probability distribution (which we call *market price distribution*), and can be represented by a *Bayesian network*. This representation might drastically reduce the computational complexity of computing/updating prices, especially when the network structure of the Bayesian network is simple. This idea was first explored by Chen et al. [5] for a class of LMSR-based combinatorial prediction markets for tournaments. They modeled the market price distribution by a Bayesian network whose graph is a balanced binary tree, and identified two types of popular securities with the following property: after any shares of any such securities are sold, the updated market price distribution can still be represented by a Bayesian network with the same structure. We call securities satisfying this property *structure-preserving* securities.

However, the approach by Chen et al. has two limitations. First, their approach only works for LMSR-based combinatorial prediction markets for tournaments, and it is not clear how to extend the results to general LMSR-based combinatorial prediction markets. Second, they only identified two types of structure-preserving securities. It is not clear whether other types of popular securities are also structure preserving.

**Our contribution.** In this paper, we significantly extend the idea in Chen at al. [5] to general LMSR-based combinatorial prediction markets. We first reveal a very natural connection between LMSR-based combinatorial prediction markets and probabilistic belief aggregation. More precisely, let Pr denote the current market price distribution, we show that for any security $F$, we can define a probability distribution $\Pr_F$ such that the updated market price distribution after selling some shares of $F$ is exactly the same as the distribution obtained by aggregating Pr and $\Pr_F$ by a well-known parameterized opinion pool function called the *logarithmic opinion pool*, or LogOP. In light of this connection and a previous work on aggregating Bayesian networks [20], we obtain a full characterization of all structure-preserving securities for any decomposable network structure (that is, in the graph, any pair of parents of any variable are connected). Notably, the main results in Chen et al. [5] are corollaries of our characterization. We then show that in order for a very basic set of securities to be structure preserving, the network structure must be decomposable, which justifies the motivation for us to focus on decomposable network structures. Finally, we briefly discuss some ideas of approximations, when the security is not structure preserving.

## 2 Preliminaries

### 2.1 LMSR-based Prediction Market

Let $\Omega = \{1, \ldots, N\}$ denote the set of outcomes of a random variable $X$. For any $i \leq N$, a security "$X = i$" means that holding each share of the security, the agent will receive \$1 from the market maker, if $X$ turns out to be $i$. In this paper, we use a *quantity vector* $\vec{q} \in \mathbb{R}^N$ to represent how many shares the market maker has sold in total for each security. That is, for each $i \leq N$, the market maker has sold $\vec{q}(i)$ shares of "$X = i$". A cost function based prediction market is characterized by a *cost function* $C : \mathbb{R}^N \to \mathbb{R}$ and an initial quantity vector $\vec{q}_0$. The price for $\epsilon$ shares of "$X = i$" is the marginal cost of incrementing the current quantity vector $\vec{q}$ by $\epsilon \vec{e}_i$ in $C$, where $\vec{e}_i$ is the $N$-dimensional vector whose $i$th component is 1 and the other components are 0. That is, if the agent wants to buy $\epsilon$ share of "$X = i$", she must pay $C(\vec{q} + \epsilon \vec{e}_i) - C(\vec{q})$ to the market maker. The instantaneous price is $\lim_{\epsilon \to 0}(C(\vec{q}+\epsilon \vec{e}_i) - C(\vec{q}))/\epsilon$, or equivalently $\partial C(\vec{q})/\partial \vec{q}(i)$. Note that price is always given in units of dollars per share, whereas cost is given in dollars.

In this paper, we focus on prediction markets with the cost function $C(\vec{q}) = b \log \sum_{i=1}^{N} e^{\vec{q}(i)/b}$, where the parameter $b$ is called the *liquidity* parameter for the market. This specific cost function corresponds to the logarithm market scoring rule (LMSR), and we call this type of prediction markets *LMSR-based prediction markets*. The following equation computes the instantaneous price $I_{\vec{q}}(i)$ for the security "$X = i$".

$$I_{\vec{q}}(i) = \frac{\partial C(\vec{q})}{\partial q_i} = \frac{e^{\vec{q}(i)/b}}{\sum_{j=1}^{N} e^{\vec{q}(j)/b}}$$

It follows that $\sum_{i \leq N} I_{\vec{q}}(i) = 1$. Therefore, we also call $I_{\vec{q}}$ the *market price distribution*.

### 2.2 Combinatorial Prediction Markets

In *combinatorial prediction markets* [4], the set of outcomes $\Omega$ has a combinatorial structure. That is, each outcome is characterized by the values of a set of variables $\mathcal{V} = \{X_1, \ldots, X_n\}$, where for each $k \leq n$, $X_k$ takes a value in a domain $\Omega_k = \{1, \ldots, l_k\}$ with $l_k \geq 2$. It follows that $\Omega = \Omega_1 \times \cdots \times \Omega_n$. In this paper, a security is represented by a logical formula $F$ over $\mathcal{V}$ in *conjunctive normal form (CNF)*.[1] That is, $F = C_1 \wedge \cdots \wedge C_K$, where for any $i \leq K$, $C_i = L_1^i \vee \cdots \vee L_{s_i}^i$, and $L_j^i$ is either $[X_k = l]$ or $\neg[X_k = l]$ (equivalently, $X_k \neq l$) for some variable $X_k \in \mathcal{V}$ and some value $l \in \Omega_k$. $C_j$ is called a *clause* and $L_j^i$ is called a *literal*. If $F$ is satisfied under the eventual true outcome (which is a valuation over $\mathcal{V}$), then the market maker should pay the agent \$1 for each share of $F$ the agent holds; otherwise the agent receives nothing for holding $F$.

---
[1] The representation of $F$ does not affect our characterization results. It only affects the computational complexity of some problems studied later in this paper.

By definition, in an LMSR-based combinatorial prediction market, the instantaneous price of $F$ is the sum of the prices of the securities that correspond to the valuations under which $F$ is satisfied. That is, $I_{\vec{q}}(F) = \sum_{\vec{v}:F(\vec{v})=1} I_{\vec{q}}(\vec{v}) = (\sum_{\vec{v}:F(\vec{v})=1} e^{\vec{q}(\vec{v})/b})/(\sum_{\vec{y}} e^{\vec{q}(\vec{y})/b})$. Chen et al. [5] have shown that the price for $\Delta b$ shares of $F$ is $b\log(e^{\Delta}I_{\vec{q}}(F) + 1 - I_{\vec{q}}(F))$. We note that computing $I_{\vec{q}}(F)$ is harder than computing marginal probabilities in $I_{\vec{q}}$, which is a well-known #P-hard problem [9].[2] However, there are many practical algorithms that compute $I_{\vec{q}}(F)$. For example, computing $I_{\vec{q}}(F)$ can be reduced to a special *weighted model counting* problem [3, 21]. In this paper, we do not focus on computing the (instantaneous) prices of securities; we will focus on the price-updating phase, that is, characterizing structure-preserving securities.

### 2.3 Bayesian Networks and Probabilistic Belief Aggregation

A Bayesian network over $\mathcal{V}$ is a compact representation for a probability distribution Pr over $\mathcal{V}$. It is composed of two parts: a direct acyclic graph (DAG) $G = (\mathcal{V}, E)$, and *conditional preference tables (CPTs)*, one for each variable $X_k \in \mathcal{V}$, which specifies the conditional probability of $X_k$ given any valuation of its parents in $G$. For any $G$ and any $X_k \in \mathcal{V}$, let $\mathbf{Pa}_G(X_k)$ denote the set of parents of $X_k$ in $G$; let $\mathbf{Ch}_G(X_k)$ denote the set of children of $X_k$ in $G$; let $\mathbf{De}_G(X_k)$ denote the set of all descendants of $X_k$ in $G$. The subscript $G$ is sometimes omitted when causing no confusion. For any valuation $\vec{v}$ of $\mathcal{V}$, $\Pr(\vec{v}) = \prod_{X \in \mathcal{V}} \Pr(\vec{v}(X)|\vec{v}(\mathbf{Pa}(X)))$, where for any $V \subseteq \mathcal{V}$, $\vec{v}(V)$ is the valuation over $V$ that agrees with $\vec{v}$. We say that Pr is *compatible* with $G$.

For any variable $X_k$, let $\mathbf{Bl}_G(X_k)$ denote the *Markov blanket* of $X_k$. That is, $\mathbf{Bl}_G(X_k) = \mathbf{Pa}(X_k) \cup \mathbf{Ch}(X_k) \cup \bigcup_{X \in \mathbf{Ch}(X_k)}(\mathbf{Pa}(X) \setminus \{X_k\})$. A DAG $G$ is *decomposable*, if for each variable $X_k \in \mathcal{V}$, there is an edge between each pair of its parents in $G$. In other words, $G$ is decomposable if for each variable $X_k$, $\mathbf{Bl}_G(X_k) = \mathbf{Pa}(X_k) \cup \mathbf{Ch}(X_k)$. A probability distribution Pr is *$G$-compatible*, denoted by Pr $\sim G$, if it can be represented by a Bayesian network whose DAG is $G$. Any $G$-compatible probability distribution Pr satisfies all *local Markov properties* in $G$: any variable $X_k$ is conditionally independent of $\mathcal{V} \setminus (\{X_k\} \cup \mathbf{Bl}_G(X_k))$ given any valuation of $\mathbf{Bl}_G(X_k)$.

Probabilistic belief aggregation has attracted a lot of interests in statistics [12, 13], and more recently in artificial intelligence [15, 18, 20]. The basic problem is to study how to aggregate the beliefs of multiple agents (represent-ed by probability distributions) to a single probability distribution. The function that takes individual distributions as inputs and outputs a single combined distribution is a called an *opinion pool function*. One of the most well-known opinion pool function is *logarithmic opinion pool* (LogOP), which is defined as follows (we only need to consider two inputs in this paper).

---

[2]In the setting of [5], computing $I_{\vec{q}}(F)$ amounts to computing a marginal probability in a tree-structured Bayesian network (see Example 1), which the well-known belief propagation algorithm [19] takes polynomial time to solve.

**Definition 1** *Let $Pr_1$ and $Pr_2$ be two probability distributions over $\Omega$. For any $\alpha_1, \alpha_2 \in \mathbb{R}$, let $\overline{Pr} = LogOP(\alpha_1 Pr_1, \alpha_2 Pr_2)$ denote the probability distribution such that for any $\vec{v} \in \Omega$, $\overline{Pr}(\vec{v}) \propto (Pr_1(\vec{v}))^{\alpha_1} \cdot (Pr_2(\vec{v}))^{\alpha_2}$.*

To better present our results, we define the following terms.

**Definition 2** *A DAG $G$ admits a security $F$, if for any $G$-compatible market price distribution Pr and any $\Delta \in \mathbb{R}$, after $\Delta b$ shares of $F$ are sold, the new market price distribution is also $G$-compatible. In this case, we also say that $F$ is* structure preserving *for $G$.*

We remark that it is possible that for a particular $G$-compatible market price distribution, after some shares of a non-structure-preserving security $F$ are sold, the new market price distribution is still $G$-compatible.

### 2.4 A Combinatorial Prediction Market for Tournaments

Chen *et al*. [5] used the following Bayesian network structure for pricing combinatorial prediction markets for tournaments of $2^m$ teams.

**Definition 3** *The tournament of $2^m$ ($m \geq 2$) teams $\{T_1, \ldots, T_{2^m}\}$ is modeled by a binary tree composed of $2^m - 1$ variables as follows. For any $1 \leq i \leq m$, let $R_i = \{X_{2^{m-i}}, \ldots, X_{2^{m-i+1}-1}\}$. For any $j \leq 2^{m-1} - 1$, $X_{2j}$ and $X_{2j+1}$ are the two children of $X_j$; $X_j$ takes a value in $\Omega_j$, which is defined recursively as follows: for any $j$ such that $2^{m-1} \leq j \leq 2^m - 1$, $\Omega_j = \{T_{2j+1-2^m}, T_{2j+2-2^m}\}$; for any $j \leq 2^{m-1} - 1$, $\Omega_j = \Omega_{2j} \cup \Omega_{2j+1}$. That is, the domain of $X_j$ is composed of all teams that can reach $X_j$, and $R_i$ is composed of all round $i$ matches. The set of all variables is $\mathcal{V}_m = R_1 \cup \ldots \cup R_m$.*

**Example 1** Figure 1 illustrates a tournament of eight teams. The domain of each variable is also shown.

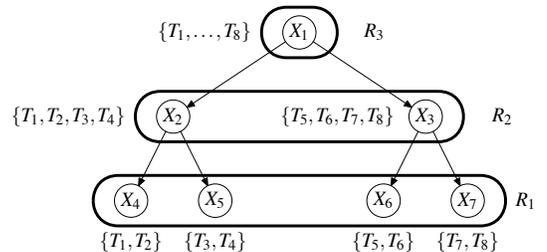

Figure 1: The Bayesian network structure for a tournament of eight teams.

In this model, not all variables in $\mathcal{V}_m$ are binary, and an outcome is a valuation of the variables in $\mathcal{V}_m$. We note that some outcomes never happen in real-life tournaments. For example, if we already have $X_1 = T_1$ (meaning that Team 1 is the winner of the whole tournament), then we cannot have $X_2 = T_2$ (which implies that Team 1 loses to Team 2 in the first round). Chen et al. [5] avoided this problem by focusing on the market price distributions that assign 0 probability to these inconsistent outcomes.

Chen et al. [5] proved the following main results: two types of popular securities are structure preserving for the DAG defined in Definition 3. The first type is composed of all securities of the form "Team $i$ wins game $X_k$" (Theorem 3.1 in Chen et al. [5]), which corresponds to the security $[X_k = T_i]$; the second type is composed of all securities of the form "Team $i_1$ and $i_2$ win games $X_{k_1}$ and $X_{k_2}$, respectively, where $X_{k_1}$ is the parent of $X_{k_2}$" (Theorem 3.2 in Chen et al. [5]), which corresponds to the security $(X_{k_1} = T_{i_1}) \wedge (X_{k_2} = T_{i_2})$.

As we will see, our results significantly extend the results by Chen et al. in two dimensions: first, we give a complete characterization of *all* structure-preserving securities, including the two types studied by Chen et al., and second, our results work for *any* combinatorial prediction market, including the one illustrated in this section.

## 3 A Natural Connection between LMSR and LogOP

In this section, we reveal a very natural connection between LMSR-based combinatorial prediction markets and probabilistic belief aggregation by LogOP. At a hight level, these two approaches (prediction markets and probabilistic belief aggregation) are close. In probabilistic belief aggregation, often, the agents' beliefs are expressed explicitly as probability distributions, so that we can directly use an opinion pool function to aggregate these distributions. On the other hand, in prediction markets, agents implicitly express part of their beliefs via their behavior in the markets (buying or selling securities), and the market price reflects the aggregated belief. Technically the connection is quite simple and intuitive: we first interpret the agents' behavior of buying a security $F$ as a probability distribution $\Pr_F$; then, we use LogOP to merge $\Pr_F$ and the current market price distribution $\Pr$, and show that the new distribution is exactly the updated market price distribution $\overline{\Pr}$ after $F$ is sold. $\Pr_F$ is defined as follows.

**Definition 4** *For any logical formula $F$, let $\Pr_F$ denote the probability distribution such that for any valuation $\vec{v}$, $\Pr_F(\vec{v}) \propto e^{F(\vec{v})}$.*

The main theorem of this section is the following.

**Theorem 1** *Let $\Delta \in \mathbb{R}$, and let $\Pr$ (respectively, $\overline{\Pr}$) denote the market price distribution before (respectively, after) an agent purchases $\Delta b$ shares of a security $F$. We have $\overline{\Pr} = LogOP(\Pr, \Delta \Pr_F)$. Moreover, if both $\Pr$ and $\Pr_F$ satisfy all local Markov properties in a DAG $G$, then $\overline{\Pr}$ also satisfies all local Markov properties in $G$.*

**Proof:** Let $\vec{q}$ denote the quantity vector corresponding to Pr. Let $\vec{1}_F$ denote the 0-1 vector over $\Omega$, where for any valuation $\vec{v}$ over $\mathcal{V}$, the $\vec{v}$ component of $\vec{1}_F$ is $F(\vec{v})$. Let $\vec{t} = \vec{q} + \Delta b \vec{1}_F$. It follows that $\overline{\Pr} = I_{\vec{t}}$. Let $\Pr^* = \text{LogOP}(\Pr, \Pr_F)$. For any pair of valuations $\vec{u}, \vec{w}$, we have $\frac{I_{\vec{t}}(\vec{u})}{I_{\vec{t}}(\vec{w})} = \frac{I_{\vec{q}}(\vec{u})}{I_{\vec{q}}(\vec{w})} \cdot \frac{e^{\Delta F(\vec{u})}}{e^{\Delta F(\vec{w})}} = \frac{\Pr(\vec{u})}{\Pr(\vec{w})} \cdot \left(\frac{\Pr_F(\vec{u})}{\Pr_F(\vec{w})}\right)^\Delta = \frac{\Pr^*(\vec{u})}{\Pr^*(\vec{w})}$. Therefore, $\overline{\Pr} = \Pr^*$.

If both $\Pr$ and $\Pr_F$ satisfy all local Markov properties in $G$, then for any variable $X_k$, any $i \leq l_k$, any valuation $\vec{v}$ of $\mathbf{Bl}_G(X_k)$, and any pair of valuations $\vec{u}, \vec{w}$ of $\mathcal{V} \setminus (\{X_k\} \cup \mathbf{Bl}_G(X_k))$, we have $\frac{\Pr(X_k = i, \vec{v}, \vec{u})}{\Pr(X_k = i, \vec{v}, \vec{w})} = \frac{\Pr_F(X_k = i, \vec{v}, \vec{u})}{\Pr_F(X_k = i, \vec{v}, \vec{w})} = 1$. Therefore, $\frac{\overline{\Pr}(X_k = i, \vec{v}, \vec{u})}{\overline{\Pr}(X_k = i, \vec{v}, \vec{w})} = \frac{\Pr(X_k = i, \vec{v}, \vec{u})}{\Pr(X_k = i, \vec{v}, \vec{w})} \cdot \left(\frac{\Pr_F(X_k = i, \vec{v}, \vec{u})}{\Pr_F(X_k = i, \vec{v}, \vec{w})}\right)^\Delta = 1$, which means that $\overline{\Pr}$ also satisfies all local Markov properties in $G$. □

We are not aware of any previous work pointing out this simple and intuitive connection. The second part of Theorem 1 states that LogOP preserves the satisfiability of local Markov properties. This observation will be useful in proving our characterization of the structure-preserving securities. We note that in Theorem 1, when $\Delta < 0$, an agent purchasing $\Delta b$ shares of $F$ from the market maker is equivalent to she selling $-\Delta b$ shares of $F$ to the market maker.

Definition 4 did not specify how $\Pr_F$ is represented (e.g., by a Bayesian network). We next provide a necessary and sufficient condition on $F$ for $\Pr_F$ to satisfy all local Markov properties in a DAG $G$.

**Definition 5** *A logical formula $F$ is compatible with a DAG $G$, if for any variable $X_k$, any valuation $\vec{v}$ of $\mathbf{Bl}_G(X_k)$, any values $i_1, i_2 \in \Omega_k$, and any pair of valuations $\vec{u}, \vec{w}$ of $V \setminus (\mathbf{Bl}_G(X_k) \cup \{X_k\})$, the following condition holds: if $F(X_k = i_1, \vec{v}, \vec{u}) \neq F(X_k = i_2, \vec{v}, \vec{u})$, then $F(X_k = i_1, \vec{v}, \vec{u}) = F(X_k = i_1, \vec{v}, \vec{w})$.*

**Example 2** *Let $G$ be the DAG in Example 1. Let $F = (X_2 = T_1) \wedge (X_5 = T_3)$. $F$ is compatible with $G$. Let $F' = (X_2 = T_1) \wedge (X_5 = T_3) \wedge (X_3 = T_8)$. $F'$ is not compatible with $G$, because $F'(X_2 = T_1, X_5 = T_3, X_3 = T_8) = 1 \neq 0 = F'(X_2 = T_1, X_5 = T_4, X_3 = T_8)$ and $F'(X_2 = T_1, X_5 = T_3, X_3 = T_8) = 1 \neq 0 = F'(X_2 = T_1, X_5 = T_3, X_3 = T_5)$. This violates the condition in Definition 5, where $k = 5$, $i_1 = T_3$, $i_2 = T_4$, $\vec{v} = [X_2 = T_1]$, $\vec{u}$ is a valuation where $X_3 = T_8$, and $\vec{w}$ is a valuation where $X_3 = T_5$.*

The condition in Definition 5 may seem unnatural at first, but in fact it has a quite intuitive explanation: $F$ is compatible with $G$ if for every variable $X_k$, the capability of changing the value of $F$ by changing the value of $X_k$ only depends on the variables in the Markov blanket of $X_k$.

**Proposition 1** *Let $G$ be a DAG. $\Pr_F$ satisfies all local Markov properties in $G$ if and only if $F$ is compatible with $G$.*

**Proof:** Suppose $\Pr_F$ satisfies all local Markov properties. We recall that for any valuation $\vec{v}$, there are only two possible values of $\Pr_F(\vec{v})$—it is either $e/\alpha$ or $1/\alpha$, where $\alpha$ is a normalizing factor. Therefore, suppose $\Pr_F(X_k = i_1, \vec{v}, \vec{u}) \neq \Pr_F(X_k = i_2, \vec{v}, \vec{u})$, then we must have $\Pr_F(X_k = i_1, \vec{v}, \vec{w}) = \Pr_F(X_k = i_1, \vec{v}, \vec{u}) \neq \Pr_F(X_k = i_2, \vec{v}, \vec{w}) = \Pr_F(X_k = i_2, \vec{v}, \vec{u})$. This corresponds to the condition in Definition 5, which proves the proposition. □

In general, satisfying all local Markov properties in $G$ is a necessary but not sufficient condition for a probability distribution to be compatible with $G$. However, it is sufficient when $G$ is decomposable, because for any variable $X_k$, $\mathbf{Bl}_G(X_k) = \mathbf{Pa}(X_k) \cup \mathbf{Ch}(X_k)$. Therefore, we immediately have the following corollaries of Theorem 1 and Proposition 1 for decomposable DAGs.

**Corollary 1** *$\Pr_F$ is compatible with a decomposable DAG $G$ if and only if $F$ is compatible with $G$.*

**Corollary 2** *Let $\Delta \in \mathbb{R}$. Let $\Pr$ (respectively, $\overline{\Pr}$) denote the market price before (respectively, after) an agent purchases $\Delta b$ shares of a security $F$. If both $\Pr$ and $F$ are compatible with a decomposable DAG $G$, then $\overline{\Pr}$ is also compatible with $G$.*

We next show that the following securities are compatible with $G$.

**Proposition 2** *Let $G$ be a decomposable DAG. Any security $F$ that only uses the variables in a clique in $G$ is compatible with $G$.*

**Proof:** Let $\mathcal{X}$ denote the set of variables used in $F$. For any $X_k \in \mathcal{X}$, because $\mathcal{X}$ is a clique, $\mathcal{X} \subseteq \mathbf{Bl}_G(X_k) = \mathbf{Pa}(X_k) \cup \mathbf{Ch}(X_k)$. Therefore, the value of any variable in $\mathcal{V} \setminus (\{X_k\} \cup \mathbf{Bl}_G(X_k))$ does not affect the value of $F$, which means that $F$ is compatible with $G$. □

In general, checking whether $F$ is compatible with a decomposable DAG is coNP-complete.

**Proposition 3** *It is coNP-complete to check whether a logical formula $F$ is compatible with a decomposable DAG $G$.*

**Proof:** We prove that the coNP-hardness holds even when all variables are binary and $G$ has no edges. For any $k \leq n$, let $\Omega_k = \{1, 2\}$. Our proof is by a reduction from the complement of the well-known NP-complete problem SAT. In a SAT instance, we are given a logical formula $H$ in CNF, and we are asked whether there exists a valuation under which $H = 1$. Without loss of generality, let $H$ be a CNF over $\{X_2, \ldots, X_n\}$ that is not a tautology. Let $\vec{u}_H$ denote an arbitrary valuation under which $H$ is false. $\vec{u}_H$ can be easily found by assigning values to make an arbitrary clause false. Let $F = (X_1 = 1) \wedge H$.

If $H$ is satisfiable, then there exists a valuation $\vec{v}_H$ under which $H$ is true. Because $F(X_1 = 1, \vec{v}_H) = 1 \neq 0 = F(X_1 = 2, \vec{v}_H)$ and $F(X_1 = 1, \vec{v}_H) = 1 \neq 0 = F(X_1 = 1, \vec{u}_H)$, $F$ is not compatible with $G$.

On the other hand, if $H$ is not satisfiable, then for any valuation $\vec{v}$ over $\{X_1, \ldots, X_n\}$, $H$ is false, which means that $F$ is always false. Therefore, $F$ is compatible with $G$. This proves the coNP-hardness of the problem.

The membership in coNP is straightforward: it is easy to verify whether the condition in Definition 5 is violated, given $k, i_1, i_2, \vec{v}, \vec{u}$, and $\vec{w}$. Therefore, it is coNP-complete to verify whether $F$ is compatible with $G$. □

## 4 Characterizing Structure-preserving Securities for Decomposable DAGs

In this section, we use the connection revealed in the last section to obtain a full characterization of structure-preserving securities for decomposable DAGs. This is the main theorem of the paper. We note that Corollary 2 implies that for $F$ to be structure preserving for a decomposable DAG $G$, it suffices for $F$ to be compatible with $G$. However, it is not immediately clear how to efficiently compute the new market price distribution. We say that a variable $X_k$ is *pivotal* in $F$, if there exists a valuation $\vec{v}$ of $V \setminus \{X_k\}$ and $i_1, i_2 \leq l_k$ such that $F(X_k = i_1, \vec{v}) \neq F(X_k = i_2, \vec{v})$.

**Theorem 2** *Let $G$ be a decomposable DAG. $F$ is structure preserving for $G$ if and only if $F$ is compatible with $G$. Moreover, after any shares of $F$ is purchased, the new market price distribution can be computed in polynomial time, where only the CPTs of the pivotal variables and the ancestors of the pivotal variables in $F$ are updated.*

**Proof:** The "only if" part is obvious. Let $\Pr_u$ denote the uniform distribution, which is compatible with $G$. Then, by Theorem 1, for any structure-preserving security $F$, the market price after $b$ shares of $F$ are sold is $\text{LogOP}(\Pr_u, \Pr_F) = \Pr_F$, which is compatible with $G$. By Proposition 1, $F$ is compatible with $G$.

Next, we prove the "if" part by presenting a polynomial-time price-updating algorithm. By Theorem 1, after $\Delta b$ shares of $F$ are purchased, the market price distribution becomes $\text{LogOP}(\Pr, \Delta \Pr_F)$. It suffices if we can directly compute the outcome of LogOP. Fortunately, when the graph is decomposable, we can use the polynomial-time algorithm devised by Pennock and Wellman (Section 3.4.1 in Pennock and Wellman [20]) to compute $\text{LogOP}(\Pr, \Delta \Pr_F)$. W.l.o.g. for any $k \leq n$, $\mathbf{Pa}(X_k) \subseteq \{X_1, \ldots, X_{k-1}\}$. The algorithm starts with the last variable $X_n$, and updates CPTs of the variables in the reverse order. Given any valuation $\vec{v}$ of $\mathbf{Pa}(X_n)$, for any $i \leq l_k$, the algorithm computes

$\overline{\Pr}(X_n = i, \vec{v})/\overline{\Pr}(X_n = 1, \vec{v})$, which uniquely determines the conditional probability $\overline{\Pr}(X_n|\vec{v})$. Then, the algorithm moves on to $X_{n-1}$, and computes the conditional probabilities in a similar way, etc. We reproduce the algorithm in Pennock and Wellman [20] for completeness of the paper (Algorithm 1). For any variable $X_k$, we let $[\mathbf{De}(X_k) = \vec{1}]$ denote the event where all descendants of $X_k$ take 1; for any $X_j \in \mathbf{De}(X_k)$, we let $[\mathbf{De}(X_k)_{-j} = \vec{1}]$ denote the event where all descendants of $X_k$, except $X_j$, take 1.

**Algorithm 1:** CompPrice

**Input**: A graph $G$, a $G$-compatible probability distribution Pr, a $G$-compatible logical formula $F$, and $\Delta \in \mathbb{R}$.
**Output**: The market price distribution $\overline{\Pr}$ after $\Delta b$ shares of $F$ are sold.

1 Compute the CPTs of $\Pr_F$.
2 $\overline{\Pr}(X_n|\mathbf{Pa}(X_n)) \propto$
$\Pr(X_n|\mathbf{Pa}(X_n))(\Pr_F(X_n|\mathbf{Pa}(X_n)))^\Delta$.
3 **for** $k = n-1$ **downto** 1 **do**
4 $\quad \overline{\Pr}(X_k|\mathbf{De}(X_k) = \vec{1}, \mathbf{Pa}(X_k)) \propto \Pr(X_k|\mathbf{De}(X_k) = \vec{1}, \mathbf{Pa}(X_k))(\Pr_F(X_k|\mathbf{De}(X_k) = \vec{1}, \mathbf{Pa}(X_k)))^\Delta$.
5 $\quad$ **for** $i = 2$ **to** $l_k$ **do**
6 $\quad\quad \dfrac{\overline{\Pr}(X_k = 1|\mathbf{Pa}(X_k))}{\overline{\Pr}(X_k = i|\mathbf{Pa}(X_k))} =$
$\dfrac{\overline{\Pr}(X_k = 1|\mathbf{De}(X_k) = \vec{1}, \mathbf{Pa}(X_k))}{\overline{\Pr}(X_k = i|\mathbf{De}(X_k) = \vec{1}, \mathbf{Pa}(X_k))} \times$
7 $\quad\quad \displaystyle\prod_{X_j \in \mathbf{De}(X_k)} \dfrac{\overline{\Pr}(X_j = 1|X_k = i, \mathbf{De}(X_k)_{-j} = \vec{1}, \mathbf{Pa}(X_k))}{\overline{\Pr}(X_j = 1|X_k = 1, \mathbf{De}(X_k)_{-j} = \vec{1}, \mathbf{Pa}(X_k))}$
8 $\quad$ **end**
9 $\quad$ Compute $\overline{\Pr}(X_k|\mathbf{Pa}(X_k))$.
10 **end**

It is not hard to see that in Algorithm 1, if a variable $X$ is neither an ancestor of any pivotal variable nor pivotal, then the CPT of $X$ is not updated. □

Theorem 2 is two-fold. On the positive side, if we can show that a security $F$ is compatible with $G$, then $F$ is very favorable for price-updating—the market price distribution after any shares of $F$ are traded is still $G$-compatible, as long as the the market price distribution before the trade is $G$-compatible. In particular, let $G$ be the graph in Definition 3, if a security $F$ is $G$-compatible, then we can ignore the assumption made in Chen et al. [5] that the inconsistent outcomes are assigned probability 0. Even though checking whether $F$ is compatible with $G$ is coNP-hard (Proposition 3), this compatibility check can be done offline.

On the negative side, requiring a security $F$ to be $G$-compatible is a strong condition—it requires that trading *any* share of $F$ does not change the network structure for *any* $G$-compatible market price distribution. It is possible that when the agents are only allowed to trade securities in some specific set (which might contain securities that are not compatible with $G$) from the very beginning, the market price distribution is always compatible with $G$. Exploring such set of securities is a hard and practical problem for future research.

We recall that in a decomposable DAG $G$, any formula $F$ that only uses the variables in a clique in $G$ is $G$-compatible (Proposition 2). Therefore, we immediately obtain the following corollary.

**Corollary 3** *Let $G$ be a decomposable DAG. Any formula $F$ that only uses the variables in a clique in $G$ is structure preserving for $G$.*

We note that the two main results in Chen et al. [5] are corollaries of Corollary 3. We recall that the Bayesian network used by Chen et al. is tree-structured (see Example 1), which is decomposable. Therefore, by Corollary 3, any security that only involves a single variable is structure preserving. In particular, for any team $i$ and any variable $X_k$, the security $[X_k = T_i]$ is structure preserving, which is Theorem 3.1 in Chen et al. [5]. For any pair of variables $X_{k_1}$ and $X_{k_2}$ where $X_{k_1}$ is the parent of $X_{k_2}$, by Corollary 3, any security that only involves $X_{k_1}$ and $X_{k_2}$ is structure preserving. In particular, for any pair of teams $T_{i_1}$ and $T_{i_2}$, the security $(X_{k_1} = T_{i_1}) \wedge (X_{k_2} = T_{i_2})$ is structure preserving, which is Theorem 3.2 in Chen et al. [5].

## 5 Network Structures That Admit a Basic Set of Securities

In this section, we characterize the Bayesian network structures that admit a basic set of securities, which justifies our focus on decomposable network structures in the last section.

To prove our theorem, we will make use of the following lemma proved by Chen et al. [5], where Pr (respectively, $\overline{\Pr}$) is the market price distribution before (respectively, after) $\Delta b$ shares of the security is purchased.

**Lemma 1 (Corollary 3.2 of [5])** *Suppose $\Delta b$ shares are purchased for event $A$, then for any events $B$ and $E$,*

$$\overline{Pr}(B|E) = Pr(B|E) \left[ \frac{e^\Delta Pr(A|BE) + Pr(\bar{A}|BE)}{e^\Delta Pr(A|E) + Pr(\bar{A}|E)} \right]$$

**Theorem 3** *Let $G$ be a DAG. If for every $k \leq n$ and every $i \leq l_k$, the security $[X_k = i]$ is structure preserving for $G$, then $G$ is decomposable.*

**Proof:** For the sake of contradiction, suppose $G$ is not decomposable. Then, there exists $k_3 \leq n$ and $X_{k_1}, X_{k_2} \in \mathbf{Pa}(X_{k_3})$ such that there is no edge between $X_{k_1}$ and $X_{k_2}$ in $G$. W.l.o.g. let $k_1 = 1, k_2 = 2, k_3 = 3$, and $X_1$ is not a descendant of $X_2$ in $G$. We define a $G$-compatible market price distribution $\Pr_*$ as follows. For any valuation $\vec{v}$ of $\mathbf{Pa}(X_3) \setminus \{X_1, X_2\}$, we let $\Pr_*(X_3 = 1|X_1 = 1, X_2 = 1, \vec{v}) = \frac{1}{4}$ and $\Pr_*(X_3 = 2|X_1 = 1, X_2 = 1, \vec{v}) = \frac{3}{4}$. For any $k' \leq n$, we let $\Pr_*(X_{k'} = 1|\mathbf{Pa}(X_{k'})) = \Pr_*(X_{k'} = 2|\mathbf{Pa}(X_{k'})) = \frac{1}{2}$, if these conditional probabilities are not defined previously. Let $\overline{\Pr}_*$ denote the market price distribution after $\Delta b$ shares of $[X_3 = 1]$ are sold to an agent. Because $[X_3 = 1]$ is structure preserving for $G$, $\overline{\Pr}_*$ is $G$-compatible. In what follows, we use Lemma 1 to prove that in $\overline{\Pr}_*$, $X_1$ and $X_2$ are not independent given any valuation $\vec{v}$ of $X_2$, which contradicts the assumption that there is no edge between $X_1$ and $X_2$ in $G$, and $X_1$ is not a descendent of $X_2$.

We note that essentially $\Pr_*$ is compatible with a DAG where there are only two edges: one from $X_1$ to $X_3$ and the other from $X_2$ to $X_3$. Therefore, in $\Pr_*$, $\mathbf{Pa}_G(X_2)$ is independent from $X_1$, $X_2$, and $X_3$. For any valuation $\vec{v}$ of $\mathbf{Pa}_G(X_2)$, we have the following calculations. For any $k = 1, 2, 3$, we let $E_k^1$ denote the valuation $X_k = 1$.

- $\Pr_*(E_3^1|\vec{v}) = \Pr_*(E_3^1) = \frac{7}{16}$.
- $\Pr_*(E_3^1|E_1^1, \vec{v}) = \Pr_*(E_3^1|E_1^1) = \frac{3}{8}$, $\Pr_*(E_3^1|E_2^1, \vec{v}) = \Pr_*(E_3^1|E_2^1) = \frac{3}{8}$.
- $\Pr_*(E_3^1|E_1^1, E_2^1, \vec{v}) = \Pr_*(E_3^1|E_1^1, E_2^1) = \frac{1}{4}$.

We calculate $\overline{\Pr}_*(E_2^1|\vec{v})$ by Lemma 1 as follows (where $A = E_3^1, B = E_2^1$, and $E = \vec{v}$).

$$\overline{\Pr}_*(E_2^1|\vec{v})$$
$$=\Pr_*(E_2^1|\vec{v}) \left[ \frac{e^\Delta \Pr_*(E_3^1|E_2^1, \vec{v}) + \Pr_*(\overline{E_3^1}|E_2^1, \vec{v})}{e^\Delta \Pr_*(E_3^1|\vec{v}) + \Pr_*(\overline{E_3^1}|\vec{v})} \right]$$
$$=\Pr_*(E_2^1) \left[ \frac{e^\Delta 6 + 10}{e^\Delta 7 + 9} \right]$$

We then calculate $\overline{\Pr}_*(E_2^1|E_1^1, \vec{v})$ by Lemma 1 as follows (where $A = E_3^1, B = E_2^1$, and $E = (E_1^1, \vec{v})$). Because $X_1$ is independent from $X_2$ in $\Pr_*$, we have $\Pr_*(E_2^1|E_1^1, \vec{v}) = \Pr_*(E_2^1|E_1^1) = \Pr_*(E_2^1)$.

$$\overline{\Pr}_*(E_2^1|E_1^1, \vec{v})$$
$$=\Pr_*(E_2^1|E_1^1, \vec{v}) \left[ \frac{e^\Delta \Pr_*(E_3^1|E_1^1, E_2^1, \vec{v}) + \Pr_*(\overline{E_3^1}|E_1^1, E_2^1, \vec{v})}{e^\Delta \Pr_*(E_3^1|E_1^1, \vec{v}) + \Pr_*(\overline{E_3^1}|E_1^1, \vec{v})} \right]$$
$$=\Pr_*(E_2^1) \left[ \frac{e^\Delta 2 + 6}{e^\Delta 3 + 5} \right]$$

For any $\Delta \neq 0$, $\frac{e^\Delta 2 + 6}{e^\Delta 3 + 5} \neq \frac{e^\Delta 6 + 10}{e^\Delta 7 + 9}$. Therefore, when $\Delta \neq 0$, $\overline{\Pr}_*(E_2^1|\vec{v}) \neq \overline{\Pr}_*(E_2^1|E_1^1, \vec{v})$, which means that given any valuation $\vec{v}$ of the parents of $X_2$ in $G$, $X_1$ and $X_2$ are not independent. However, because $\overline{\Pr}_*$ is $G$-compatible and $X_1$ is not a descendant of $X_2$, we must have that $X_1$ and $X_2$ are independent given any valuation of the parents of $X_2$, which is a contradiction. Therefore, $G$ is decomposable. □

It follows from Theorem 1 and Corollary 3 that for any variable $X_k$ and any value $i \leq l_k$, the security $[X_k = i]$ is structure preserving for any decomposable DAG over $\mathcal{V}$, which leads to the following theorem.

**Theorem 4** *A DAG $G$ is decomposable if and only if for every $k \leq n$ and every $i \leq l_k$, the security $[X_k = i]$ is structure preserving for $G$.*

## 6 Approximate Price Updating

For a security $F$, being compatible with a sparse decomposable DAG is a strong condition. Therefore, in practice it is important to study how to approximately update the market price distribution when a security $F$ is not compatible with $G$, while keeping the Bayesian network structure $G$ the same. Intuitively there are at least two ways to do so. The first is to approximate $F$ by a $G$-compatible logical formula $F'$ that is closest to $F$ in terms of Hamming distance. Then, after $F$ is sold, we update the market price distribution as if $F'$ is sold. We do not pursue this approach in this paper.

In this section, we propose an approximation in light of the connection we revealed in Section 3. We recall that $F$ corresponds to a distribution $\Pr_F$ (Definition 4), and subsequently price-updating in LMSR-based markets corresponds to aggregating $\Pr$ and $\Pr_F$ by LogOP. Therefore, we seek for a probability distribution $\Pr_F^*$ that is compatible with $G$, and is as close to $\Pr_F$ as possible. A natural metric that measures the closeness between two probability distributions is the *KL divergence* (also known as *relative entropy*). For any pair of probability distributions $\Pr$ and $\Pr'$, $\mathrm{KL}(\Pr, \Pr') = \sum_{\vec{v}} \Pr(\vec{v}) \log(\Pr'(\vec{v})/\Pr(\vec{v}))$. We note that KL divergence is not a distance, because it is not commutative, that is, usually $\mathrm{KL}(\Pr, \Pr') \neq \mathrm{KL}(\Pr', \Pr)$. Then, after $\Delta b$ shares of $F$ is sold, we update the market price distribution by using Algorithm 1 to compute $\mathrm{LogOP}(\Pr, \Delta \Pr_F^*)$, where $\Pr$ is the market price distribution before $F$ is sold.

Given $\Pr_F$, finding $\Pr_F^*$ that minimizes its KL divergence from $\Pr_F$ is a classical problem in Bayesian network learning, for which there is a conceptually simple and intuitive solution, defined as follows.

**Definition 6** *For any logical formula $F$, let $Pr_F^*$ denote the $G$-compatible distribution where for any variable $X_k$, $Pr_F^*(X_k|\mathbf{Pa}_G(X_k)) = Pr_F(X_k|\mathbf{Pa}_G(X_k))$.*

That is, in $\Pr_F^*$, the CPT entry of $X_k$ given any valuation $\vec{v}$ of the parents of $X_k$ in $G$ is exactly the same as the conditional probability $\Pr_F(X_k|\vec{v})$.

**Theorem 5 (Theorem 17.2 in Darwiche [10])**

$$Pr_F^* = \arg\min_{Pr \sim G} KL(Pr_F, Pr)$$

When $F$ is not $G$-compatible, $\Pr_F(X_k|\mathbf{Pa}_G(X_k))$ might be hard to compute. The following proposition simplifies the computation, whose proof directly follows the definition of $\Pr_F$ and Bayes' rule, and therefore is omitted.

**Proposition 4** *For any variable $X_k$, any set of variables $Y$ such that $X_k \notin Y$, any valuation $\vec{y}$ of $Y$, and any $i \leq l_k$,*

*we have:*

$$Pr_F(X_k = i|\vec{y}) =$$

$$\frac{(e-1)g_F(X_k = i, \vec{y}) + \prod_{X_l \notin (Y \cup \{X_k\})} |\Omega_l|}{(e-1)g_F(\vec{y}) + \prod_{X_l \notin Y} |\Omega_l|}$$

*Here for any set of variable $Z$ and any valuation $\vec{z}$ of $Z$, $g_F(\vec{z}) = |\{\vec{v} : F(\vec{v}) = 1 \text{ and } \vec{v}(Z) = \vec{z}\}|$, where $\vec{v}(Z)$ is the valuation over $Z$ that agrees with $\vec{v}$.*

Therefore, computing $\Pr_F(X_k = i|\vec{y})$ amounts to computing $g_F(X_k = i, \vec{y})$ and $g_F(\vec{y})$. These are standard model counting problems, which are #P-complete. However, there are many practical algorithms for computing their exact values as well as approximations [14].

## 7 Future Work

A number of questions remain for future research. Can we design efficient (randomized/approximation) algorithms to compute the price of a security when the market price distribution is represented by a sparse decomposable Bayesian network?[3] How can we evaluate our approximation technique in real-life combinatorial prediction markets? Are there any better approximation methods? Can we find a natural set of (not necessarily structure-preserving) securities such that the market price distribution is always $G$-compatible, if only the securities in this set are traded? More generally, we believe that the connection between prediction markets and probabilistic belief aggregation will shed some light on designing better combinatorial prediction markets.

## Acknowledgements

Lirong Xia acknowledges a James B. Duke Fellowship and Vincent Conitzer's NSF CAREER 0953756 and IIS-0812113, and an Alfred P. Sloan fellowship for support. We thank Yiling Chen and all UAI-11 reviewers for helpful suggestions and comments.

---

[3] We recall that computing the prices of the two types of securities in the setting of Chen et al. [5] is easy, because it corresponds to computing marginal probabilities in tree-structured Bayesian networks.